\documentclass[12pt]{article}
\usepackage{amsmath,amssymb,graphicx,mathrsfs,slashed}
\usepackage[usenames, dvipsnames]{color}
\usepackage{tabularx}
\usepackage[%
verbose,
	colorlinks=true,
naturalnames=true,
linkcolor=blue,
	citecolor=black,
]{hyperref}
\newcommand{\be}{\begin{equation}}
        \newcommand{\ee}{\end{equation}}
\newcommand{\bea}{\begin{eqnarray}}
        \newcommand{\eea}{\end{eqnarray}}

\def\({\left(} \def\){\right)}

\begin{document}
%%%%%%%%
\title{\vspace{-1.8in}
%\vspace{3mm}
%\vspace{0.3cm}
{Quantum State of Black-Holes Out of Equilibrium }}

\author{\large Yotam Sherf ${}^{(1)}$
\\
\vspace{-.5in}  \vbox{
\begin{center}
	$^{\textrm{\normalsize
			(1)\ Department of Physics, Ben-Gurion University,
			Beer-Sheva 84105, Israel}}$
	\\ \small %\hspace{-2.in}
sherfyo@post.bgu.ac.il
\end{center}
}}
\date{}
\maketitle
%%%
\begin{abstract}
%\abstract

\vspace{0.2cm} We study the properties of black-holes (BHs) that are out of equilibrium about the Hartle-Hawking (HH) vacuum state. We show how gravitational perturbations excite the vacuum state, thus making it a superposition of states, which eventually leads to additional excitations. We examine the vacuum state structure in the presence of time-dependent gravitational perturbation. As a result, we determine the vacuum state evolution and calculate the semi-classical modifications to the particle occupation number of the emitted particles. We demonstrate that the quantum emission in BHs that are far away from equilibrium is comparable and even larger than the Hawking radiation. 
\end{abstract}

\newpage
\renewcommand{\baselinestretch}{1.5}\normalsize

\begin{subequations}
\renewcommand{\theequation}{\theparentequation.\arabic{equation}}
\section{Introduction}
The vacuum state properties in curved spacetime and especially for Schwarz-schild BHs were thoroughly studied at \cite{Davies:1976ei,Unruh:1976db,D,DeWitt:1975ys,Candelas:1980zt}. There it is shown in the semi-classical approach, in which the matter fields are quantized about a fixed curved background. That the vacuum expectation value (VEV) of the quantized matter fields is non-vanishing, rather the short distance behavior of the fields product in the stress energy momentum (SEM) tensor makes the VEV divergent. In general, the finite value of the SEM tensor VEV is attained when considering appropriate regularization techniques. The method considers the appearance of higher-order curvature correction terms in short distances \cite{Candelas:1980zt,B,Calzetta:1986ey,Christensen:1976vb,Christensen:1978yd,Page:1982fm}, this eventually leads to a cancellation of the inherent infinities and a finite VEV. This indicates that in curved backgrounds, the creation of particles from the vacuum is possible as well as other interesting phenomena, which implies that the vacuum state properties in curved space are non-trivial. In this paper, we investigate its properties for the spherically symmetric Schwarzschild BHs that are out of equilibrium.

  Here, as a continuation of the authors previous work \cite{Brustein:2019twi}. We elaborate on some of the calculations and apply them to examine the modifications to the HH vacuum state in BHs that are out of equilibrium. The BHs are perturbed by a gravitational external source where in general, the source for the perturbation is irrelevant. Here, we show how gravitational perturbations that are coupled to the quantized matter fields, modify the vacuum state thus leading to additional excitations and modified particle occupation number. Furthermore, we examine the quantum states of these excitations and identify their structure in the wave-packet basis. Then, we demonstrate how large deviations from the  Schwarzschild background leads to an amplified Hawking radiation. 

The paper is organized as follows, in the first part we consider the presence of an external weak gravitational field, then, in the interaction picture, we demonstrate how the vacuum state time evolution is dictated by the external field. In the second part, we examine the time-dependent vacuum state and interpret it in terms of excited energy states and also identify the explicit terms that govern the vacuum excitations. These terms are found to be divergent, we avoid the divergence by using the known result of the renormalized SEM tensor from the literature. Then, we establish the theoretical framework that describes the geometry of BHs that are out of equilibrium. Next, we calculate the vacuum state time evolution and its associated energy which eventually enables the calculation of the modified particle occupation number. In the last part, we estimate the modifications to the Hawking radiation for the case of BHs that are strongly perturbed. We show the large background deviations lead to a quantum emission that is comparable or even larger than the Hawking radiation.
In the Appendix, for completeness, we calculate directly the renormalized value of the divergent terms that govern the vacuum excitations. 
\end{subequations}

\begin{subequations}
\renewcommand{\theequation}{\theparentequation.\arabic{equation}}

\section{Time evolution in perturbed backgrounds}\label{subsection1.1}
Here, we discuss the consequences of gravitational perturbations about the HH vacuum state of stationary BHs. Particularly interesting is the vacuum state time evolution in the presence of external perturbations and its implications about the matter field excitations. For simplicity, we consider the case of a minimally-coupled massless scalar field in vacuum whose matter action is given by\footnote{The generalization of the other spin fields is given below.}  
\begin{equation}
S_M = \frac{1}{2}\int d^4x \sqrt{-g} g_{ab}\nabla^a \phi \nabla ^b \phi \label{GN1}~.
\end{equation}
We consider gravitational perturbations about the background Schwarzschild metric. 
%The perturbation is finite in a sense that it is only present during the time interval $ t\in\{t_i,t_f\}$. 
Then  $\bar{g}_{ab}~=~g_{ab}+h_{ab}$ where the perturbation is assumed to be small $|h_{ab}|\ll1$, which implies that the expansion of the matter fields action to first order is given by
\begin{gather}
	\delta S(t)_M~=~\dfrac{1}{2}\int d^4x \sqrt{-g}H_{ab}  \nabla^a {\phi} \nabla ^b {\phi}~,\label{SM1}
\end{gather} 
where $H_{ab}=h_{ab}-\frac{1}{2}g_{ab}h^c_{~c}$, and also  $|h_{ab}|\ll1$. The application of the external time-dependent perturbation in some early time divided the quantum states of the fields into two distinct states, namely the \textit{in} and \textit{out} states. In general, in curved spacetime even in the absence of perturbations the \textit{in} and \textit{out} states are distinct such that particles are produced. Therefore it is obvious that in the presence of external perturbations the number of particle excitation increases, as we demonstrates below. 

 To realize how gravitational perturbations modify the HH vacuum state, we find the amplitude of the time-dependent \textit{in} state to the time-independent \textit{out} state, namely the $\langle	\text{out},0|0, \text{in} \rangle_t$. Then, in the interaction picture, where the explicit time-dependence of the \textit{in} state is assigned to the corresponding time-dependent operator $\delta \hat{S}(t)_M$, the time promotion operator is identified  as $\hat{U}=e^{i\delta \hat{S}(t)_M}$, so the \textit{in} state time evolution is given by 
 \begin{gather}
 	|0, \text{in} \rangle_t~=~e^{i\delta \hat{S}(t)_M}|0 ,\text{in} \rangle~.
 	\label{TPO1}
 	 	 \end{gather}	 	
 	 	In general, the expression for the time promotion operator appears naturally when considering the vacuum persistence amplitude (VPA) in the presence of external source term $Z[J]=\langle \text{out},0 |0,\text{in}\rangle_J$. Where in the gravitational context, the analog of the external source term $J$ is the gravitational perturbation $h_{ab}$, so $Z[h]=\langle\text{out},0 |0,\text{in}\rangle_h$. We assume a quasi-static perturbation such that it is adiabatically switched on at some early time in the past \textit{in} region\footnote{Below we show explicitly that the time scale associated with the gravitational perturbation is significantly larger than the Schwarzschild time scale, which agrees with the quasi-static assumption.}. Then the VPA can be written as \cite{B,Calzetta:1986ey} 
 	 	\begin{gather}
\langle\text{out},0 |0,\text{in}\rangle_h~=~\langle	\text{out},0|T~\text{exp}\left[{\frac{i}{2}\int dt\int d^3x {\sqrt{-g}}H_{ab}\nabla^a {\hat{\phi}} \nabla ^b {\hat{\phi}}}\right]|0, \text{in} \rangle~,
\label{GTC}
 	 	\end{gather}
 where $T$ stands for temporal order.  Following the definition of the effective action $\langle	\text{out},0|0, \text{in} \rangle=e^{i\mathcal{W}}$, the generating functional becomes	 	
 	 \begin{equation}
 	 \begin{split}
 	Z[h]~&=~{e^{i\mathcal{W}}}~	\langle e^{\frac{i}{2}\int d^4x {\sqrt{-g}}H_{ab}\nabla^a {\hat{\phi}} \nabla ^b {\hat{\phi}}}\rangle ~\\&\approx~~e^{i\mathcal{W}}~e^{\frac{i}{2}\int d^4x {\sqrt{-g}}H_{ab}\langle \nabla^a {\hat{\phi}} \nabla ^b {\hat{\phi}}\rangle_{\text{ren}}}~.
 	 \end{split}
 	 \end{equation}
 	Where in the second line we approximate according to $|H_{ab}|\ll 1$.

 	  Then, by expanding the time promotion operator of the \textit{in} state to first order in the perturbation   we obtain
 \begin{gather}
 \langle	\text{out},0|0, \text{in} \rangle_h ~\approx ~	e^{i\mathcal{W}} ~\left(1+{\frac{i}{2}\int d^4x {\sqrt{-g}}H_{ab}\langle \nabla^a {\hat{\phi}} \nabla ^b {\hat{\phi}}\rangle_{\text{ren}}}~\right).
 \label{OI}
 \end{gather}
 Where the \textit{ren} indicates that the renormalized VEV has to be considered. The reason is that in curved spacetime, a straightforward calculation of the VEV  is divergent and therefore must be renormalized. In general, the divergent terms in the matter action are geometric and can be absorbed into the gravitational action, which includes the appearance of the higher-order curvature correction terms \cite{Candelas:1980zt,Christensen:1976vb,Christensen:1978yd,Page:1982fm}. This eventually leads to the cancellation of the inherent infinities and to a finite VEV that is labeled as $\langle \nabla^a {\hat{\phi}} \nabla ^b {\hat{\phi}}\rangle_\text{ren}$.   %$\langle	\text{in},0| \nabla^a {\hat{\phi}} \nabla ^b {\hat{\phi}}|0, \text{in} \rangle_{\text{ren}}=\langle \nabla^a {\hat{\phi}} \nabla ^b {\hat{\phi}}\rangle_\text{ren}$

 Where for the HH vacuum state the effective action is defined by the partition function as
 \begin{equation}
 \begin{split}
 i\mathcal{W}~&=~\text{ln}{Z[0]}\\&=~-\sum_j \text{ln}\left( 1-e^{-\omega_j/T_H}\right)~.	\end{split}\label{28}
 \end{equation}
  The particle occupation number is given by 
 \begin{gather}
 	N_j~=~\dfrac{1}{e^{\omega_j/T_H}-1}\label{PD}~.
 \end{gather}
 It is now clear that the first term in Eq.~(\ref{OI}) describes a thermal state with the Hawking temperature $T_H^{~-1}=8\pi M$ and a black-body emission spectrum, whereas second term is a modification to the Hawking term induced by the external perturbation.
 
 \subsection{Vacuum state excitations} \label{VSS}
 To proceed, we show explicitly how the external gravitational source modify the VPA. First, we expand the field operators in terms of their corresponding creation and annihilation operators,
 \begin{equation}
 \hat{\phi}_{\text{in}}~=~\sum_j(\hat{a}_ju_j+\hat{a}^\dagger_j u^*_j),
 \label{61}
 \end{equation}
  the \textit{in} vacuum is associated with an early time, unperturbed state that is defined by $ \hat{a}_i|0, \text{in} \rangle=0$, and also by the modes $u_j$ which denote the corresponding  positive-energy ingoing propagating mode solutions in the \textit{in} region%with respect to v
 . Then, the expansion of Eq.~(\ref{OI}) in the field operators reads,
 \begin{gather}
 \begin{split}
 \langle	\text{out},0|0, \text{in} \rangle_h~\approx&~
 \left[1-\dfrac{i}{2}\sum_{i,j}\delta_{ij}\int d^4x\sqrt{-g}H_{ab}\nabla^au_j\nabla^b u^*_i\right]_{\text{ren}}\langle	\text{out},0|0, \text{in} \rangle\\-
 &~\left[\dfrac{i}{2}\sum_{i,j}\int d^4x \sqrt{-g}H_{ab}\nabla^au_j^*\nabla^b u^*_i\right]_{\text{ren}}\langle\text{out},0|1_i,1_j, \text{in}\rangle
 \label{1.18}~.
 \end{split}
 \end{gather}
 Obviously, the transition amplitude is a sum of three distinct terms that stem from different origins.
 The first term is the unperturbed partition function $e^{i\mathcal{W}}$, which in the case of the HH vacuum state describes a BH in thermal equilibrium with a black-body
 spectrum. This term is being recovered in the absence of perturbations $H_{ab}=0$. The additional two contributions, emerge in the presence of external perturbation, are essentially modification to the Hawking radiation.  The first contribution to the amplitude takes the form of the HH vacuum state augmented by a perturbative dependent factor and a second contribution that describes an initial non-vacuum state composed of an excited pair particle state.

 Next, to reveal the composition of Eq.~(\ref{1.18}), in particular the excited pair particles state which is not proportional to the effective action. We decompose the excited  \textit{in} eigenstates in terms of the vacuum  \textit{out} eigenstates. Where the \textit{in} to \textit{out} transformation is given by the Bogoliubov coefficients, then following \cite{D,DeWitt:1975ys,B} the decomposition takes the form 
 \begin{equation}
 \begin{split}
 &\langle	\text{out},0|1_{l_1},1_{l_2},\dots\dots,1_{l_{m-1}},l_{m},\text{in} \rangle~=\\&~-i^{m/2+1}\langle	\text{out},0|0, \text{in} \rangle 	 \sum_{\{l_1 \dots l_m\}}\sum_k\beta_{kl_{2}}\alpha^{-1}_{kl_{1}} \cdots\cdots \beta_{kl_{m}}\alpha^{-1}_{kl_{m-1}}~~~~~\text{{{m even}}},
 \label{115}
 \end{split}
 \end{equation}
 and vanishes for {$m$ odd}.
 The indices $ \{l_1 \dots l_m\} $ represent a sum over all distinct permutation of $ \{l_1 \dots l_m\} $. The Bogoliubov coefficients $\beta_{kl_{i+1}}\alpha^{-1}_{kl_{i}}$ defines the representation of particles in the \textit{in} state with momenta $l_{i+1},l_{i}$ to the vacuum \textit{out} state. To demonstrate that, we consider the expansion of the field operators in terms of the positive-energy mode solutions in the \textit{out} region 
 \begin{gather}
 \hat{\phi}_{out}~=~\sum_j(\hat{b}_j\bar{u}_j+\hat{b}^\dagger_j\bar{u}^*_j)~,
 \label{62}
 \end{gather} 
 where now the \textit{out} vacuum is defined by $\hat{b}_i|0,\text{out} \rangle=0$, and the  $\bar{u}_i$ are its analogous positive frequency outgoing propagating modes. Then, the Bogoliubov transformation def\hspace{0.02cm}ines the expansion of the  \textit{in} and \textit{out} modes and their associated operators by
 \begin{gather}
 u_i~=~\sum_j\left(\alpha^*_{ji}\bar{u}_j-\beta_{ji}\bar{u}^*_j\right),~~~~~~~u_j~=~\sum_i\left(\alpha_{ji}{u}_i+\beta_{ji}{u}^*_i\right), \label{BT1}\\
 ~	a_i~=~\sum_j\left(\alpha_{ji}\hat{b}_j+\beta^*_{ji}\hat{b}^{\dagger}_j\right),~~~~~~~\hat{b}_i~=~\sum_j\left(\alpha^*_{ji}{a}_i-\beta^*_{ji}{a^{\dagger}_i}\right).
 \label{BT}
 \end{gather} 
 
 For the specific case of a pair particles state as in Eq.~(\ref{1.18}), we consider particles with momenta  $l_1, l_2$, so the expansion of Eq.~(\ref{115}) takes the simple form
 \begin{gather}
 \langle	\text{out},0|1_{i},1_{j},\text{in} \rangle~=~\langle	\text{out},0|0, \text{in} \rangle\sum_k \beta_{kj}\alpha^{~-1}_{ki}~.
 \label{141}
 \end{gather}	
 Applying these relations, Eq.~(\ref{1.18}) can be written as 
 \begin{gather}
 \langle	\text{out},0|0, \text{in} \rangle_t~=
 \left[	1-i\left(\sum_{m,n}\delta_{mn}A_{mn}+\sum_{i,j,k}B_{ijk} \right)
 \right] e^{i\mathcal{W}}~,
 \label{MS}
 \end{gather}
 and the renormalized coefficients $A_{mn}, B_{ijk}$ are defined as
 \begin{align}
 A_{mn}~=~\left[\frac{1}{2}\int d^4x\sqrt{-g}~H_{ab}\nabla^au_m\nabla^b u^*_n~\right]_{\text{ren}}~,\label{1.21}\\
 B_{ijk}~=~\left[\frac{1}{2}\int d^4x\sqrt{-g}~H_{ab}\nabla^au_j^*\nabla^b u^*_i{\beta_{kj}}{\alpha_{ki}^{-1}}\right]_{\text{ren}}~.
 \label{1.20}
 \end{align}
 Noticing that in the absence of perturbations $A_{mn}, B_{ijk}=0$,  the transition amplitude reduces to $\langle	\text{out},0|0, \text{in} \rangle~=~e^{i\mathcal{W}}~.$
 As previously mentioned, $ren$ implies that these terms are divergent and their infinites need to be discarded through an appropriate renormalization procedure, so eventually only finite pieces of the integrals are involved. To show that, we explicitly solve the EOM for a massless free scalar field in Schwarzschild spacetime $\Box \hat{\phi}=0$ and find its asymptotic mode function solution $u_j$, \cite{Hawking:1974sw} 
 \begin{gather}
 u_{jlm}~=\dfrac{Y_{lm}}{\sqrt{4\pi \omega_j}r}e^{-i\omega_jv}~, 
 \label{UIN}
 \end{gather}
 as well as their corresponding Bogoliubov coefficients Eq.~(\ref{BT1})
 \begin{gather}
 \alpha_{ki}~=~t_ke^{i(\omega_k-\omega_i)v_0}\Gamma\left(1-i\dfrac{\omega_k}{\kappa}\right)\left(\dfrac{\omega_i}{\omega_k}\right)(-i\omega_i)^{1-i\frac{\omega_k}{\kappa}}e^{\frac{\pi \omega_k}{2\kappa}}~.
 \label{1.27}
 \end{gather}
 Where $\kappa=1/4M$ and $t_k$ is the transmission coefficient of the mode $\omega_k$ that is determined by the matching conditions about the potential barrier \cite{Page:1976df}. $v_0$ is the horizon formation time according to the collapsing shell model \cite{Hawking:1974sw}. In this case, the perturbation is switched on for a time interval $\Delta v=v_f-v_i$ at some early enough time $v_f<v_o$. Then in the presence of perturbation, additional radiation is being emitted, which eventually arrives to infinity.
 
  Another important identity is the relation between the coefficients $\alpha_{ik}$ and $\beta_{ik}$ which is given by
 \begin{gather}
 \beta_{kj}~=~-i\alpha_{k,-j}~,~~~~~~~~~~~~~~~~~~~~~~~~~~\alpha_{ki}~=~i\beta_{k,-i}
 \label{220}
 \end{gather}
 With these relations, it is possible to show that $A_{mn}\sim \omega$ so its mode space summation leads to a quadratic divergence $\sim \omega^2$, whereas the divergence of the $B_{ijk}$ term is not obvious and its expression must be further simplified in order to be noticed (for a detailed demonstration see Appendix \ref{app}).
 As a consequence, we have to subtract these infinities by introducing an appropriate renormalization procedure,  where here the proper technique for discarding the infinities in the mode space representation is the adiabatic mode sum regularization \cite{Bunch:1979uk,Bunch:1980vc}.

To understand the behavior of the higher-order terms Eq.~(\ref{MS}) and especially that of the $B_{ijk}$ term, we examine the transformation factor ${\beta_{kj}}{\alpha_{ki}}^{-1}$ in its mode space representation. This term is also considered as the annihilation amplitude as implied by Eq.~(\ref{141}).  To illustrate this, we first assume that the matrix $\alpha_{ki}$ is diagonal, as in the Hawking case, this assumption relies on a direct calculation of the off-diagonal terms in the $\alpha_{ki}$ matrix. As demonstrated in \cite{Alberte:2015cxa}, where it is shown that the leading order terms of the off-diagonal matrix elements $\alpha_{ki}$ are essentially correction terms to the power one in the perturbation strength. Therefore, the inclusion of the off-diagonal correction terms in the $\alpha_{ki}$ matrix leads to a negligible second-order correction in the expression of the $B_{ijk}$ integral. Hence, we consider that both $\alpha_{ki}$ and $\beta_{kj}$ are diagonal. Then, in the Fourier basis the annihilation term takes the form
\begin{equation}
\begin{split}
	\dfrac{\beta_{kj}}{\alpha_{ki}}~&=~-e^{i(\omega_i+\omega_j)v_0}\left(\dfrac{\omega_j}{\omega_i}\right)^{-1+i\frac{\omega_k}{\kappa}}e^{-\frac{\pi \omega_k}{\kappa}}\\&=~-
e^{i(\omega_i+\omega_j)v_0}\left(\dfrac{\omega_i}{\omega_j}\right)e^{i\frac{\omega_k}{\kappa}\text{ln}\frac{\omega_j}{\omega_i}}e^{-\frac{\pi \omega_k}{\kappa}}~.
\label{F1}
\end{split}
\end{equation}
The exponential dependence suggests that the majority of particles in the excited \textit{in} state are annihilated by the corresponding particles in the vacuum \textit{out} state with energy levels $\omega_k\lesssim \kappa$. Whereas the contribution of the energetic particles with $\omega_k > \kappa$ is attenuated by the exponent term. 
To see that, we approximate the $\omega_k$ summation in the high frequency regime $\omega_k > \kappa$ 
 \begin{gather}
 \sum_{\omega_k \gtrsim \kappa}e^{i\frac{\omega_k}{\kappa}\text{ln}\frac{\omega_j}{\omega_i}}e^{-\frac{\pi \omega_k}{\kappa}}~~\approx~\sum_{\omega_k \gtrsim \kappa} e^{-\frac{\pi \omega_k}{\kappa}}~,
 \end{gather}
 on the other hand, for the regime $\omega_k<\kappa$ the sum becomes 
   \begin{gather}
  \sum_{\omega_k < \kappa}e^{i\frac{\omega_k}{\kappa}\text{ln}\frac{\omega_j}{\omega_i}}e^{-\frac{\pi \omega_k}{\kappa}}~~\approx~\left(\pi+i\text{ln}\left(\dfrac{\omega_i}{\omega_j}\right)\right)\sum_{\omega_k < \kappa}\frac{\omega_k}{\kappa}+\mathcal{O}\left(\dfrac{\omega_k}{\kappa}\right)^2~,
  \end{gather}
  which constitutes most of the contribution.

 To proceed, the calculation of the sum is performed by converting the sum into integral. According to \cite{Hawking:1974sw,Alberte:2015cxa}, to do that we sum over wave packets that were emitted during the presence of the external perturbation, such that the duration of the perturbation $\Delta t$ is long enough in comparison with the wave packet mode width. 
Then, in the wave packet basis the sum is converted into an integral according to
 \begin{gather}
 \sum_j\rightarrow \dfrac{\Delta t}{2\pi}\int d \omega_j~.
 \label{STI}
 \end{gather}
 Therefore
 \begin{equation}
 \begin{split}
 	\sum_k\dfrac{\beta_{kj}}{\alpha_{ki}}~&=~-\dfrac{\Delta t}{2\pi}e^{i(\omega_i+\omega_j)v_0}\left(\dfrac{\omega_i}{\omega_j}\right)\int 
 	e^{i\frac{\omega_k}{\kappa}\text{ln}\frac{\omega_j}{\omega_i}}e^{\frac{\pi \omega_k}{\kappa}}d \omega_k~\\&=~-\dfrac{\Delta t}{2\pi}e^{i(\omega_i+\omega_j)v_0}\left(\dfrac{\omega_i}{\omega_j}\right)\dfrac{\kappa}{\pi}\left(\dfrac{1}{1-\frac{i}{\pi}\text{ln}\frac{\omega_i}{\omega_j}}\right)~.
 	\label{boa}
 	\end{split}
 \end{equation}
In the following section we demonstrate the calculation of the integral coefficients, the
rest of the derivation of the amplitude coefficients is given in the Appendix \ref{app}.
 
  \subsection{Integral coefficients}
  Next, in order to calculate the expression Eq.~(\ref{MS}) we express Eq.~(\ref{GTC}) in terms of the stress-energy-momentum tensor (SEM), recalling that the matter fields variation is related to SEM tensor by
 \begin{gather}
 {\delta S_M}~=~\dfrac{\sqrt{-g}}{2} h_{ab}T^{ab},
 \label{SM}
 \end{gather}
and by substituting this expression back into Eq.~(\ref{OI}) we get
 \begin{gather}
 \langle	\text{out},0|0, \text{in} \rangle_h ~\approx ~	e^{i\mathcal{W}} ~\left(1+\frac{i}{2}\int d^4x {\sqrt{-g}}h_{ab}\langle T^{ab}\rangle_{\text{ren}}~\right).
 \label{GTC1}
 \end{gather}
Thus, the perturbation dependent terms in Eq.~(\ref{MS}) can be written as
 \begin{gather}	
 -~
 \frac{1}{2}\int d^4x {\sqrt{-g}}h_{ab}\langle T^{ab}\rangle_{\text{ren}}~=~
\left[\sum_{m,n}\delta_{mn}A_{mn}+\sum_{i,j,k}B_{ijk} \right]_{\text{ren}}~.
 \label{RQ}
 \end{gather}
Now we consider the VEV of the RSEM (renormalized SEM) tensor, about the Hartle-Hawking vacuum state which accounts
 to a situation where a BH is immersed in a thermal bath with an infinite sea of black body radiation. The RSEM tenor is attributed to all fields species, in particular for the above mentioned minimally coupled scalar as well as the other spin fields.
 So, the finite contribution of the perturbation dependent term can be easily calculated by considering the appropriate VEV of the RSEM tensor that is given at  \cite{Page:1982fm}.
 We stress that the RSEM (renormalized SEM) tensor is attributed to all field species. Especially for the above mentioned massless scalar field we have
 \begin{gather}
 T^{ab}~=~\dfrac{1}{2}\nabla^a\phi\nabla^b\phi-\dfrac{1}{4}g^{ab}\nabla^c\phi\nabla_c\phi~.
 \label{2.88}
 \end{gather}
\end{subequations}

\begin{subequations}
	\renewcommand{\theequation}{\theparentequation.\arabic{equation}}
 \section{{Gravitational perturbations}}\label{BD}
 In the following section, we establish the theoretical framework that describes the geometry of a tidally deformed BH. Here we consider the case of a BH that is slightly perturbed, the specific details of the mechanism are unimportant. The source for the perturbation can be either due to the inspiral of a BH with other remote compact objects or by the post-merger event where the BH is relaxing to equilibrium or by a small injection of energy. The important issue is that the gravitational perturbations that deform the BH horizon and excites the vacuum state eventually increase in the matter field excitations.

 \subsection{Geometry of Black-holes out of equilibrium}\label{GBD}
 
 Here we are particularly interested in the realistic scenario of perturbations that are generated during the  inspiral of BH with an external equal mass remote compact object. However, since the evaluation of the integrals Eqs.~(\ref{1.21}),(\ref{1.20}) involves the entire Schwarzschild spacetime, including the coordinate divergence in $R_S=2M$, we must require a well-defined perturbation at this point. A unique description for this deformed geometry was formulated by  Poisson \cite{Poisson:2004cw,Poisson:2009qj,Taylor:2008xy,Poisson:2018qqd}, who found by applying a specific coordinates gauge, the so called \textit{horizon locking gauge}, a description for the metric perturbation that is finite about the horizon. Before writing down the explicit expression for the metric of a BH that is tidally deformed, we briefly review the analogous description of inspiralling binaries.

The idea is that the spacetime of a non-rotating BH of mass $M$ is perturbed by a remote equal mass moving object. The companions relative distance is $b\gg R_S$, so the binary period time scale $T\approx \sqrt{b^3/M}$ is significantly larger than the typical Schwarzschild time scale $T\gg R_S$, which implies that the two objects are weakly interacting the therefore their orbits are slowly varying and well described by a circular orbit with an orbital velocity $v=\sqrt{M/b}$.   
 We also stress that modifications to the background Schwarzschild metric as a result of deformation in the horizon are significant at the BH vicinity, which is defined by $R_S<r<\sqrt{bR_S}$ \cite{Thorne:1980ru}. Another relevant scale is the associated length scale of the characteristic time scale $T$, referred as the radius of curvature $\mathcal{R}=T$ that satisfy $r_{max}\ll\mathcal{R}$, so for a radial distance  $r<r_{max}$ from the BH center we get $r/\mathcal{R}\ll1$. For a more detailed description we refer the readers to \cite{Vega:2011ue}. Here we label the maximal radius where the metric perturbation is valid by $r_{max}=\sqrt{bR_S }$, where for $r>r_{max}$ deformations in the BH horizon induce negligible corrections to the background metric that are higher order in the parameter $r/\mathcal{R}$ \cite{Poisson:2004cw}.
 
 The perturbed metric is considered about the non-rotating Schwarzschild background in the outgoing Eddington-Finkelstein (EF) coordinates whose line element is given by
  \begin{gather}
 ds^2~=~-f(r)du^2-2dudr+r^2d\Omega^2
 \end{gather}
 where $f(r)=1-2M/r$. 
The metric perturbation is given as an expansion in the dimensionless parameter $r/\mathcal{R}\ll1$ that is kept fixed. Here we only list the first non-vanishing term in the expansion, which as explained in \cite{Brustein:2019twi,Poisson:2004cw} is the contribution of the quadrupole moment $l=2$ and constitutes the dominant contribution to the GW emission. The higher-order terms are the octupole moment $l=3$ of order $\sim ({r}/{\mathcal{R}})^{3/2}$ and the $l=4$ hexadecapole moment $\sim ({r}/{\mathcal{R}})^{2}$. We also find that due the spherical symmetry of $\langle T^{ab}\rangle_{\text{ren}}$ and the fact that the azimuthal components of $h_{ab}$ are traceless, the relevant term that is obtained by the contraction  $h_{ab}\langle T^{ab}\rangle_{\text{ren}}$ in Eq.~(\ref{GTC1}) involves only the $uu$ components.  Then, in the vicinity of the BH horizon, the metric perturbation takes the form
\begin{equation}
h_{uu}~=~-\left(1-\dfrac{2M}{r}\right)^2\dfrac{r^2}{\mathcal{R}^2}
\label{MP}
\end{equation}
\end{subequations}

\begin{subequations}
\renewcommand{\theequation}{\theparentequation.\arabic{equation}}
  \section{Calculations}
  Following the geometric description that is given above, our main purpose in this section is to evaluate the modifications to the particle occupation numbers for BHs that are out of equilibrium. In order to do that we first specify the vacuum state time evolution Eq.~(\ref{GTC1}) as given below.
\subsection{Vacuum state time evolution}\label{VN}
To determine the time evolution ot the vacuum state we recall Eq.~(\ref{GTC}) and compare it to the general   time evolution expression $|0, \text{in} \rangle_t=e^{-{i}\int_0^t dt'\langle \Delta \hat{ E} \rangle}|0, \text{in} \rangle~$. In this form we identify the energy difference $\langle \Delta{ E} \rangle$
\begin{gather}
\langle \Delta { E} \rangle~=~-\frac{1}{2}\int d^3x {\sqrt{-g}}h_{ab}\langle { T}^{ab}\rangle_{\text{ren}}~.
\label{ddd}
\end{gather}
The excess energy of the time-dependent vacuum state with respect to the stationary initial vacuum state is supplied by the external gravitational perturbation. It is considered as the total energy gained by the matter fields as a result of the gravity-matter coupling  $h_{ab}\langle T^{ab}\rangle$.

Because of the spherical symmetry of RSEM tensor and since the angular components of $h_{ab}$ (the $h_{AB}$ part) are traceless, the only relevant components of the RSEM tensor for spin zero particles in the HH vacuum state are \cite{Page:1982fm}
\begin{gather}
\langle	T^{t}_{~t}\rangle_{\text{ren}}^{{0}}=\dfrac{\pi^2}{30}T^4\left(24\left(\dfrac{2M}{r}\right)^6-\dfrac{1-(4-6M/r)^2(2M/r)^6}{(1-2M/r)^2}\right)~,\label{RSEM1}\\
\langle	T^{r}_{~r}\rangle_{\text{ren}}^{{0}}=\dfrac{\pi^2}{90}T^4\left(24\left(\dfrac{2M}{r}\right)^6+\dfrac{1-(4-6M/r)^2(2M/r)^6}{(1-2M/r)^2}\right)~. \label{RSEM2}
\end{gather}
Transforming the outgoing EF coordinates to Schwarzschild coordinates, we find that $h_{uu}=h_{tt}=h_{rr}f(r)^{-2}$, and the summation yields $h_{ab}T^{ab}=h_{uu}\left(T^r_{~r}-T^t_{~t}\right)f(r)^{-1}$.

For the integration limits, we point out that the RSEM tensor is defined in the exterior BH region, which sets the integral lower bound $r>R_S$. The upper bound is determined according to the validity region of the deformed BH metric Eq.~(\ref{MP}). Therefore the major contribution to the integral Eq.~(\ref{ddd}) emerges from the BH vicinity $R_S<r< \sqrt{b R_S}$ and in terms of the orbital velocity $\sqrt{bR_S}=R_S/v$. We stress that as a result of the post-Newtonian expansion in the vicinity of the BH, the contribution to $\langle \Delta E \rangle$ from the far regions $r>R_S/v$ induce higher order corrections to the background metric than that of Eq.~(\ref{MP}) and therefore can be neglected. We calculate the excess energy that is prominently governed by the quadrupole $l=2$ modes as dictated by the metric perturbation Eq.~(\ref{MP}), we label it as $\langle  \Delta E \rangle^s_{2}$ where $s$ denotes the spin of the different matter fields that are expressed by the RSEM tensor. Then,  for spin zero scalar particles\footnote{We use the approximation  $\langle  \Delta E \rangle^s_{2}\approx \langle  \Delta E \rangle^s$ since most of the\vspace{-0.2cm} energy stems from the $l=2$ quadrupole perturbation Eq.~({\ref{MP}}).}
\begin{equation}
\begin{split}
\langle  \Delta E \rangle^{0}~&=~-\frac{1}{2}\int_{R_S}^{ R_S/v} d^3x \sqrt{-g}h_{ab}\langle T^{ab}\rangle_{\text{ren}}~
\\&
\approx~	\dfrac{T_H}{900} \left(v+\dfrac{5}{4}v^2+\mathcal{O}(v^3)\right)~,
\label{2.E}
\end{split}\end{equation}
 where for spin half neutrinos \footnote{All flavors are assumed to be massless} and spin one photons we obtain
\begin{align}
\langle  \Delta { E} \rangle^{1/2}~&\approx~\dfrac{7}{2}\langle  \Delta { E} \rangle^0~,\\\langle  \Delta { E} \rangle^{1}~&\approx ~13\langle  \Delta { E} \rangle^0~.
\label{2.N}
\end{align}
For the total energy gained by the vacuum state as a result of horizon deformation we get $
\sum_{s}\langle  \Delta E \rangle^s\approx\frac{1}{50}T_H v$.
 \subsection{Modified particle occupation number}
 	Next, to evaluate the particle occupation number, we calculate the perturbed partition function.
 The calculation of the partition function is carried out by recalling that in equilibrium it is in a thermal state with a characteristic temperature $T_H$ and it is given by Eq.~(\ref{28}).
 Where in the presence of perturbations, the partition function is approximately thermal and characterized by a black-body spectrum and a modified temperature $\widetilde{T}$. According to Sec.~\ref{GBD}, the thermalization occurs due to the long duration of the perturbation $\Delta t\sim T$ and since it is slowly varying in time such that the system reaches thermal equilibrium. Then, the partition function and its associated states are thermalized with a modified characteristic temperature $\widetilde{T}$. Under these assumptions, the modified partition function takes the form of Eq.~(\ref{28}) by the replacement $T_H \rightarrow \widetilde{T}$.
 Then, from the relation between the effective action and the particle number occupation Eq.~(\ref{PD}) we obtain
 \begin{equation}
 \begin{split}
  \text{Im}\widetilde{\mathcal{W}}~&\approx~\sum_l\text{ln}(1+\widetilde{N}_l)\\
 &=-\sum_l\text{ln}\left(1-e^{- \omega_l/\widetilde{T}}\right)~.
 \label{MEF}
 \end{split}
 \end{equation}
 Where $\mathcal{\widetilde{W}}$ is the modified effective action, which is evaluated about the Euclidean time by 
 \begin{gather}
 {\mathcal{\widetilde{W}}}~=~\mathcal{W}+\frac{1}{2}{\int d^4x {\sqrt{-g}}h_{ab}\langle T^{ab}\rangle_{\text{ren}}}~,\label{53}
 \end{gather} and $\widetilde{N}_l$ is the modified particle occupation number which is approximately given by a Planck spectrum with a characteristic temperature $\widetilde{T}$
 \begin{gather}
 \widetilde{N}_l~\approx~\dfrac{1}{e^{\omega_l/\widetilde{T}}-1}~.
 \label{MPD}
 \end{gather}
To evaluate $\widetilde{T}$ we convert the sum in Eq.~(\ref{MEF}) to an integral according to Eq.~(\ref{STI}), this leads to 
 \begin{gather}
-\sum_j\text{ln}(1-e^{-\omega_j/\widetilde{T}})~=~-\Delta t
\int_0^{\infty}d\omega_j\text{ln}(1-e^{-\omega_j/\widetilde{T}})~=~\Delta t \dfrac{\pi^2 \widetilde{T}}{6}~,
\label{TM}
 \end{gather}
 where the non-perturbed effective action is given in terms of the Hawking temperature by $\text{Im}\mathcal{W}=\Delta t T_H\pi^2/6$, which brings Eq.~(\ref{53}) to the form
 \begin{gather}
 	\Delta t \dfrac{\pi \widetilde{T}}{12}~=~\Delta t \dfrac{\pi {T}}{12}+\dfrac{\Delta t}{2}{\int d^3x {\sqrt{-g}}h_{ab}\langle T^{ab}\rangle_{\text{ren}}}~.
 \end{gather}
  Then, we combine the results of $\mathcal{\widetilde{W}}$ and $\mathcal{W}$ and also  Eqs.~(\ref{2.E})-(\ref{2.N}), which eventually yields the modified temperature of the perturbed HH vacuum state for the different kinds of matter fields
 \begin{gather}
\widetilde{T}~\approx~T_H\left(1+\dfrac{6}{25\pi}v\right)
\label{TM1}
\end{gather}

To understand how significant are the modifications to the particle occupation number, we examine the ratio $\Delta N_l/N_l$ where $\Delta N_l=\widetilde{N}_l-N_l$ and expand it to first order in $|\Delta T|=T_H\frac{6v}{25\pi}$ which for non-relativistic velocities $\Delta T/T_H\ll1$ reads
\begin{equation}
\begin{split}
	\dfrac{\Delta{N}_l}{{N}_l}~&\approx~\dfrac{\omega_l/T_H}{1-e^{-\omega_l/T_H}}\dfrac{6}{25\pi}v~\\&~=~\dfrac{1}{1-e^{-\omega_l/T_H}}\dfrac{\omega_l\Delta T}{T_H^2}~.
	\end{split}
\end{equation}
It is clear that for the low frequency regime $\omega<< T_H$ we get $\Delta N_l/N_l\approx \Delta T/T_H\ll1$ so the gravitational perturbation induce negligible corrections.
For the  regime $\omega\approx T_H$ we get $\Delta N_l/N_l\approx v/8$.
An increment in the particle occupation number implies that $\Delta N_l/N_l>1$ and it is being displayed for the frequency regime $\omega \Delta T\gtrsim T_H^2$, or in term of the velocity  ${\omega_l}
\frac{6v}{25\pi}\gtrsim{T_H}$, therefore gravitational deformations induce considerable modifications to $\widetilde{N}_l$.

\end{subequations}

\begin{subequations}
\renewcommand{\theequation}{\theparentequation.\arabic{equation}}
\section{Amplified Hawking radiation}
Here, we estimate the modifications to the Hawking radiation when the BH is far away from equilibrium, as in the case of BHs collision. We show that when BHs collide, the quantum emission becomes comparable or larger than the Hawking radiation. In this case, since there is no analytic expression for the geometry of colliding BHs, it is  only possible to estimate the amplified quantum radiation.
\subsection{General arguments}
We consider the emission of GWs during the merger event of a BH with another massive object. We will not restrict the emitted GWs amplitude to be small, so non-linear effects are present. The only requirement is that the GWs are evaluated far enough from the emitted source $r\gg R_S$, so the waves propagate in empty space and the EOM are source free. 
Hence, following the description \cite{Price:1994pm}, in the far-field approximation $r\gg R_S$ the dominant GW amplitude takes the form $h\sim \frac{1}{r}f(M,R_S,\chi,\Delta R_S)$, where the function $f(M,R_S,\chi,\Delta R_S)$ encodes the information from the collision event, such as the spin $\chi$ and  $\Delta R_S$ that measures the radial deviations of the deformed to the stationary horizon.

We emphasize that in order to demonstrate the amplification of the quantum emission such that it is not a correction to the Hawking radiation, there is no need to consider the contribution from the higher-order gravitational effects. Rather at $r\gg R_S$, it is sufficient to consider the contribution from the Spherically symmetric Schwarzschild background. Moreover, although considerable contributions are ignored, the subtracted remaining contribution from the Schwarzschild background leads to an amplified quantum emission.

To estimate the magnitude of the GW at an arbitrary distance $r$ from the source, we recall that at a distance $D$ from the source a detector on earth experience a relative length deviation of $h(D)\sim\Delta L/L\sim 10^{-21}$. In addition to $h\sim\frac{1}{D}f(M,R_S,\chi)$, we conclude that the wave amplitude at a distance $r$ scales as 
\begin{gather}
h(r)\sim\dfrac{\Delta L}{L}\dfrac{D}{r}
\end{gather}
To evaluate this, for a relative deviation $\Delta L/L\sim 10^{-21}$ and a distance $D\sim 1$Gpc the amplitude at $r\sim10R_S$ becomes $h(10R_S)\sim 1$.

Next, in order to estimate the magnitude of the emitted quantum radiation and simplify the calculations we follow Sec.~\ref{subsection1.1}, where the massless scalar field action is $S_M = \frac{1}{2}\int d^4x \sqrt{-g} g_{ab}\nabla^a {\phi} \nabla ^b {\phi}$.
 In this case, the background geometry is not explicit and can be described as an unspecified colliding BH geometry labeled by $\left({g}_{ab}\right)_C$, an isolated Schwarzschild BH $\left(g_{ab}\right)_S$ and a traveling GW about the background, so the background metric becomes
\begin{gather}
g_{ab}=\left(g_{ab}\right)_S+\left(g_{ab}\right)_C+h_{ab}
\label{5.2}
\end{gather} 
 In similar to Sec.~\ref{subsection1.1}, the modified metric Eq.~(\ref{5.2}) brings a gravity-matter coupling term into the matter action. Here we are particularity interested in the coupling of the kinetic term to the traveling GW metric  $h_{ab}\left[\langle\nabla^a \hat{\phi} \nabla ^b \hat{\phi}\rangle_{\text{ren}}\right]_S$, where $S$ indicates that the renormalized VEV is evaluated about the stationary Schwarzschild background.
Next, in order to evaluate the measure in the action integral we use the identity Det$(g)\geq$  Det$(g_S)$+Det$(g_C)$, so Det$(g_S)<$ Det$(g)$. Also, since $\left[\langle\nabla^a \hat{\phi} \nabla ^b \hat{\phi}\rangle_{\text{ren}}\right]_S<\langle\nabla^a \hat{\phi} \nabla ^b \hat{\phi}\rangle_{\text{ren}}$, we ignore the non-analytical contributions to the matter action and define the subtracted matter action 
\begin{gather}
\widetilde{S}_M=~ \frac{1}{2}\int d^4x \sqrt{-g_S} h_{ab}\left[\langle\nabla^a \hat{\phi} \nabla ^b \hat{\phi}\rangle_{\text{ren}}\right]_S~,
\end{gather} 
which satisfies $S_M> \widetilde{S}_M$. 
For convenience we approximate the kinetic term by using the Schwarzschild SEM tensor Eq.~(\ref{2.88})  
\begin{equation}
\begin{split}
\left[\langle\nabla^a \hat{\phi} \nabla ^b \hat{\phi}\rangle_{\text{ren}}\right]_S~&=~\left[\langle T^{ab}-\frac{1}{2}\left(g^{ab}\right)_S T^c_{~c}\rangle _{\text{ren}}\right]_S~.\\&\Big|_{r\gg R_S}\approx~\left[\langle T^{ab}\rangle _{\text{ren}}\right]_S
\label{STM1}
\end{split}
\end{equation}
where the last equality stems from the SEM tensor trace anomaly $\left[\langle T^{c}_{~c}\rangle_{\text{ren}}\right]_S\sim \frac{1}{r^6}$ that provides a negligible contribution to the emitted radiation at the far-field approximation \cite{Page:1982fm}. Then,
following Sec.~\ref{subsection1.1} we obtain the associated subtracted energy difference 
\begin{gather}
\langle \Delta\widetilde{ E} \rangle ~=~ -\frac{1}{2}\int d^3x \sqrt{-g_S} h_{ab}\left[\langle T^{ab}\rangle_{\text{ren}}\right]_S~.
\end{gather}

To demonstrate the amplified radiation, we mention that the additional energy gained by the vacuum $\langle \Delta\widetilde{ E} \rangle$ has an energy flux contribution that radiates to infinity. This emitted radiation is given by the flux components of the SEM tensor (the null components) which are given in EF coordinates by $T^{uu}$. Therefore the additional luminosity at a distance $r$ is given by \cite{Brustein:2019twi}
\begin{gather}
\Delta \widetilde{L}~~= ~-\dfrac{1}{2}\int d\Omega r^2  |h(r)_{uu}| \langle T^{uu}\rangle_{\text{ren}}~,
\end{gather}
then, at a distance $r=10R_S$ from the merged objects $h(10R_S)\sim 1$, the additional luminosity becomes 
\begin{gather}
\Delta \widetilde{L}\big|_{r\rightarrow 10R_S}\sim \int d\Omega r^2  \langle T^{uu}\rangle_{\text{ren}} \sim L_H
\end{gather}
The conclusion is that the additional luminosity at $r=10R_S$ is of order of the Hawking luminosity. Obviously  for $r<10R_S$ the additional luminosity becomes larger than the Hawking luminosity since in this region $|h(r)|>1$.
The conclusion is that BHs that are far away from equilibrium induce large background modifications that eventually generates amplified quantum emission.

  \section{Summary and Discussion}
In this paper, we examined the HH vacuum state properties of BHs that are out of equilibrium. Following the semi-classical approach, we identify the explicit expression for the vacuum state time evolution in the presence of time-dependent gravitational perturbations. We find that terms that determine the vacuum state time evolution are divergent and can be calculated by introducing the renormalized SEM tensor\footnote{The complete derivation of the divergent terms from Eqs.~(\ref{1.20}),(\ref{1.21}) is given in the Appendix \ref{app}.}. Then, in Sec.~\ref{VSS} we analyze the vacuum state structure in the mode representation in particular for the high-frequency limit as well as for the other limits, which are governed by the ratio ${\beta_{kj}}/{\alpha_{ki}}$.
Then, we established the theoretical framework for the gravitational perturbation, we adopted the \textit{horizon locking gauge} in which the perturbation is finite about the BH horizon. Following the geometric description,  we determined the vacuum state time evolution Eq.~(\ref{ddd}) and its associated energy difference that is supplied by the external perturbation. Then, we evaluate the modifications to the particle occupation number from the definition of the partition function and the effective action.  There we assumed that the partition function is approximately thermal with a black-body
spectrum and a different characteristic temperature $\widetilde{T}$.
It is important to mention that the resulted modifications to the temperature and the particle occupation number, are not quantum corrections that stems from additional fluctuations, rather they should be viewed as a semi-classical modifications that are reflected in the gravity-matter coupling of the gravitational perturbation and the vacuum fluctuation of the matter fields in curved background $h_{ab}\langle\nabla^a \hat{\phi} \nabla^b \hat{\phi}\rangle$. 
In the last section, we estimated the modifications to the Hawking luminosity when the BH is far from equilibrium, as in the case of BHs collision.  We show that when large curvature modifications are induced to the unperturbed background, the quantum emission becomes comparable or even larger than the Hawking emission. The results were estimated in the far-field approximation and by ignoring the non-linear effect as well as additional fluctuation terms. The results suggest that in the vicinity of the merged objects, where the geometric description is non-analytic and non-linear effects are significant, the curvature modification to the background becomes dominant, therefore the quantum emission is larger or comparable to the Hawking radiation. As a consequence, it would be interesting to give an exact estimation of the additional quantum emission and its observability odds. Since, as demonstrated in Sec.~\ref{VN} all particle species are emitted, in particular photons and neutrinos. Then, perhaps in the extreme case of a super-massive BHs collision, the emission of quantum origin is detectable. 
\section*{Acknowledgments}

I thank Ramy Brustein and Yoav Zigdon for discussions and comments.
The research of YS was supported by the Israel Science Foundation grant no. 1294/16.  The research of YS was supported by the BGU Hi-Tech scholarship.

\end{subequations}\vspace{1.5cm}
 \begin{subequations}
 	\renewcommand{\theequation}{\theparentequation.\arabic{equation}}
\appendix
\LARGE{\textbf{Appendix}}\normalsize\setcounter{equation}{0}
\section{Renormalized Integral coefficients}\label{app}
\renewcommand{\theequation}{\Alph{section}.\arabic{equation}}
In this appendix we calculate the integral coefficients that are given in
\begin{gather}
\langle	\text{out},0|0, \text{in} \rangle_t~=
\left[	1-i\left(\sum_{m,n}\delta_{mn}A_{mn}+\sum_{i,j,k}B_{ijk} \right)
\right] e^{i\mathcal{W}}
\label{AMS1}
\end{gather}
where the coefficients $A_{mn}, B_{ijk}$ are defined by \begin{align}
A_{mn}~=~\left[\frac{1}{2}\int d^4x\sqrt{-g}~H_{ab}\nabla^au_m\nabla^b u^*_n~\right]_{\text{ren}}\delta
_{mn},\label{a1.21}\\
B_{ijk}~=~\left[\frac{1}{2}\int d^4x\sqrt{-g}~H_{ab}\nabla^au_j^*\nabla^b u^*_i{\beta_{kj}}{\alpha_{ki}^{-1}}\right]_{\text{ren}}~~.
\label{a1.20}
\end{align}
Here we perform an explicit calculation of $A_{mn}, B_{ijk}$ in the horizon locking gauge.
For simplicity, we consider a radially outgoing wave in the out region with zero angular momentum $l,m=0$. So its mode function is given in  by 
\begin{gather}
u_{j}~=\dfrac{e^{i\omega_ju}}{\sqrt{ \omega_j}}f(r)~.
\end{gather}
Where $f(r)$ is the radial dependence of the mode function\footnote{ In general, $f(r)$ depends on the external potential barrier and can be derived by solving the corresponding wave equation. Its form in the vicinity of the horizon and in infinity is given from Eq.~(\ref{UIN}) by $f(r)=\frac{1}{\sqrt{4\pi}r}$.}. It is unimportant to know the exact values of $f(r)$ for the calculation of the coefficients, as shown below.

Then, from the definition of the horizon locking gauge \cite{Poisson:2004cw}, the only metric component that calculations is $H_{uu}=h_{uu}(2-g_{uu})$.

Now, $A_{mm}$ can be expressed as
\begin{equation}
\begin{split}
A_{mn}~&=~\frac{1}{2}\int d^4x\sqrt{-g}~H_{ab}\nabla^au_m\nabla^b u^*_n~
,\label{A1.21}\\
&=~\dfrac{1}{2}\int d^4x\sqrt{-g}H_{uu}\nabla^uu_m\nabla^u u^*_n~\\~&=~
\frac{1}{2}\int dud^3x\sqrt{-g}H_{uu}f(r)\sqrt{{\omega_m}\omega_n} e^{i(\omega_m-\omega_n)u}
\end{split}
\end{equation}
To demonstrate the divergence we recall that in the wave packet basis the summation is converted to an integral according to Eq.~(\ref{STI}), then $A_{mn}$ becomes
\begin{gather}
A_{mm}~=~\frac{\Delta t}{4\pi}\int d^3x\sqrt{-g}H_{uu}f(r){\omega_m}
\end{gather}
It is now clear that $A_{mm}\sim \omega^2$ and must  be renormalized. This divergence is also present in the $B_{ijk}$ term, as we show below
\begin{align}
\nonumber B_{ijk}~&=~\frac{1}{2}\int d^4x\sqrt{-g} H_{ab}\nabla^au_j^*\nabla^b u^*_i\dfrac{\beta_{kj}}{\alpha_{ki}}~\\&=~\nonumber
\frac{1}{2}\int d^4x\sqrt{-g}H_{uu}\nabla^uu_j^*\nabla^u u^*_i\dfrac{\beta_{kj}}{\alpha_{ki}}~\\&=
-\frac{1}{2}\int dud^3x\sqrt{-g}H_{uu}f(r)\sqrt{\omega_i\omega_j}{e^{i(\omega_i+\omega_j)u}} \dfrac{\beta_{kj}}{\alpha_{ki}}~\\&=~\nonumber
\frac{i}{2}\int_0^{\infty}d\tilde{u}e^{i(\omega_i+\omega_j)\tilde{u}}\int d^3x\sqrt{-g}H_{uu}f(r)\sqrt{\omega_i\omega_j}\dfrac{\omega_j}{\omega_i}\left(\dfrac{\omega_j}{\omega_i}\right)^{2+i\frac{\omega_k}{\kappa}}e^{\frac{-\pi \omega_k}{\kappa}}~\\&=~\nonumber
\frac{1}{2}\delta(\omega_i+\omega_j)\int d^3x\sqrt{-g}H_{uu}f(r)\sqrt{\omega_i\omega_j}\left(\dfrac{\omega_j}{\omega_i}\right)^{2+i\frac{\omega_k}{\kappa}}e^{\frac{-\pi \omega_k}{\kappa}}~=
\\\nonumber&=~\frac{\Delta t}{4\pi}\int d^3x\sqrt{-g}H_{uu}f(r){\omega_i}\left(-1\right)^{2+i\frac{\omega_k}{\kappa}}e^{\frac{-\pi \omega_k}{\kappa}}~.
\label{A1.2}
\end{align}
Where in the fourth line, according to \cite{Hawking:1974sw} we define $\tilde{u}=u+v_0$  and the ratio $ \beta_{kj}/\alpha_{ki}$ is taken from Eqs.~(\ref{1.27})(\ref{220}). For the $\sum_{ijk}B_{ijk}$ we get
\begin{align}
\sum_k B_{jjk}~&=~\dfrac{\Delta t^2}{8\pi^2}\int d\omega_k\left(-1\right)^{2+i\frac{\omega_k}{\kappa}}e^{\frac{-\pi \omega_k}{\kappa}}\times\frac{i}{2}\int d^3x\sqrt{-g}H_{uu}f(r){\omega_i}~
\\\nonumber&=~
\dfrac{\kappa \Delta t }{4\pi^2}\times  \sum _j A_{jj}~
\end{align}
Finally, we substitute the coefficients in Eq.~(\ref{AMS1}) and obtain
\begin{gather}
\langle	\text{out},0|0, \text{in} \rangle_t~=
\left[	1-i\left(1+\dfrac{\kappa \Delta t}{4 \pi^2}\right)\sum_{m}A_{mm} 
\right] 
\end{gather}
Now, it is possible to calculate the renormalized sum $\left[\sum_m A_{mm}\right]_{\text{ren}}$ from the relations Eqs.~(\ref{GTC1}),(\ref{RQ})
\begin{gather}
\dfrac{\Delta t}{2 \pi}\left(1+\dfrac{\kappa \Delta t}{8 \pi^2}\right) \left[\sum_m A_{mm}\right]_{\text{ren}}
	 ~=~{\frac{1}{2}\int d^4x {\sqrt{-g}}h_{ab}\langle T^{ab}\rangle_{\text{ren}}}
\end{gather}
which eventually by applying Eqs.~(\ref{ddd}),(\ref{TM}) and also from $T_H=\frac{\kappa}{2\pi}$, becomes
\begin{gather}
	\left[\sum_m A_{mm}\right]_{\text{ren}}~\approx~ \dfrac{T_H \Delta t }{1+\frac{T_H \Delta t}{4 \pi}}\dfrac{v}{900}
\end{gather}
\end{subequations}

%%%%%%%%%%%%%%%%%%%%%%%%%%%%%%%%%%%%%%%%%%%%%%%%%%%%%%%%%%%%%%%%%%%%%%%%%%%%%%%%%%%%%%%%%%%%
%%%%%%%%%%%%%%%%%%%%%%%%%%%%%%%%%%%%%%%%%%%%%%%%%%%%%%%%%%%%%%%%%%%%%%%%%%%%%%%%%%%%%%%%%%%%

\begin{thebibliography}{99}
		{\small	
		
					%\cite{Davies:1976ei}
		\bibitem{Davies:1976ei} 
		P.~C.~W.~Davies, S.~A.~Fulling and W.~G.~Unruh,
		``Energy Momentum Tensor Near an Evaporating Black Hole,''
		Phys.\ Rev.\ D {\bf 13}, 2720 (1976).
		%	doi:10.1103/PhysRevD.13.2720
		%%CITATION = doi:10.1103/PhysRevD.13.2720;%%
		%246 citations counted in INSPIRE as of 15 Jan 2019
		
		
		
			
			%\cite{Unruh:1976db}
			\bibitem{Unruh:1976db} 
			W.~G.~Unruh,
			``Notes on black hole evaporation,''
			Phys.\ Rev.\ D {\bf 14}, 870 (1976).
		%	doi:10.1103/PhysRevD.14.870
			%%CITATION = doi:10.1103/PhysRevD.14.870;%%
			%2861 citations counted in INSPIRE as of 19 Jul 2019
			
				
			%\cite{DeWitt:2003pm}
			\bibitem{D}
			B.~S.~DeWitt,
			``The global approach to quantum field theory. Vol. 1, 2,''
			Int.\ Ser.\ Monogr.\ Phys.\  {\bf 114}, 1 (2003).
			%%CITATION = IMPHA,114,1;%%
			%83 citations counted in INSPIRE as of 22 Mar 2018
			%\cite{Poisson:2004cw}
			
				%\cite{DeWitt:1975ys}
			\bibitem{DeWitt:1975ys} 
			B.~S.~DeWitt,
			``Quantum Field Theory in Curved Space-Time,''
			Phys.\ Rept.\  {\bf 19}, 295 (1975).
			%doi:10.1016/0370-1573(75)90051-4
			%%CITATION = doi:10.1016/0370-1573(75)90051-4;%%
			%1025 citations counted in INSPIRE as of 07 Feb 2019
			
				%\cite{Candelas:1980zt}
			\bibitem{Candelas:1980zt}
			P.~Candelas,
			``Vacuum Polarization in Schwarzschild Space-Time,''
			Phys.\ Rev.\ D {\bf 21}, 2185 (1980).
			%doi:10.1103/PhysRevD.21.2185
			%%CITATION = doi:10.1103/PhysRevD.21.2185;%%
			%277 citations counted in INSPIRE as of 13 Nov 2018
			
					%\cite{Birrell:1982ix}
			\bibitem{B}
			N.~D.~Birrell and P.~C.~W.~Davies,
			``Quantum Fields in Curved Space,''
			%doi:10.1017/CBO9780511622632
			%%CITATION = doi:10.1017/CBO9780511622632;%%
			%1671 citations counted in INSPIRE as of 08 Apr 2018
			
			%\cite{Calzetta:1986ey}
			\bibitem{Calzetta:1986ey} 
			E.~Calzetta and B.~L.~Hu,
			``Closed Time Path Functional Formalism in Curved Space-Time: Application to Cosmological Back Reaction Problems,''
			Phys.\ Rev.\ D {\bf 35}, 495 (1987).
		%	doi:10.1103/PhysRevD.35.495
			
				%\cite{Christensen:1976vb}
			\bibitem{Christensen:1976vb}
			S.~M.~Christensen,
			``Vacuum Expectation Value of the Stress Tensor in an Arbitrary Curved Background: The Covariant Point Separation Method,''
			Phys.\ Rev.\ D {\bf 14}, 2490 (1976).
			%doi:10.1103/PhysRevD.14.2490
			%%CITATION = doi:10.1103/PhysRevD.14.2490;%%
			%411 citations counted in INSPIRE as of 18 Jun 2018
			
			%\cite{Christensen:1978yd}
			\bibitem{Christensen:1978yd}
			S.~M.~Christensen,
			``Regularization, Renormalization, and Covariant Geodesic Point Separation,''
			Phys.\ Rev.\ D {\bf 17}, 946 (1978).
			%doi:10.1103/PhysRevD.17.946
			%%CITATION = doi:10.1103/PhysRevD.17.946;%%
			%272 citations counted in INSPIRE as of 23 Jun 2018
			
				%\cite{Page:1976df}
		\bibitem{Page:1976df} 
		D.~N.~Page,
		``Particle Emission Rates from a Black Hole: Massless Particles from an Uncharged, Nonrotating Hole,''
		Phys.\ Rev.\ D {\bf 13}, 198 (1976).
		%doi:10.1103/PhysRevD.13.198
		%%CITATION = doi:10.1103/PhysRevD.13.198;%%
		%695 citations counted in INSPIRE as of 15 Jan 2019
			
			%\cite{Page:1982fm}
			\bibitem{Page:1982fm}
			D.~N.~Page,
			``Thermal Stress Tensors in Static Einstein Spaces,''
			Phys.\ Rev.\ D {\bf 25}, 1499 (1982).
			%doi:10.1103/PhysRevD.25.1499
			%%CITATION = doi:10.1103/PhysRevD.25.1499;%%
			%257 citations counted in INSPIRE as of 22 Aug 2018
			
			
			%\cite{Brustein:2019twi}
			\bibitem{Brustein:2019twi} 
			R.~Brustein and Y.~Sherf,
			``Emission Channels from Perturbed Quantum Black Holes,''
			arXiv:1902.08449 [hep-th].
			%%CITATION = ARXIV:1902.08449;%%

	
%\cite{Hawking:1974sw}
\bibitem{Hawking:1974sw}
S.~W.~Hawking,
``Particle Creation by Black Holes,''
Commun.\ Math.\ Phys.\  {\bf 43}, 199 (1975)
Erratum: [Commun.\ Math.\ Phys.\  {\bf 46}, 206 (1976)].
%doi:10.1007/BF02345020, 10.1007/BF01608497
%%CITATION = doi:10.1007/BF02345020, 10.1007/BF01608497;%%
%7348 citations counted in INSPIRE as of 13 Nov 2018
			
						%\cite{Bunch:1979uk}
			\bibitem{Bunch:1979uk}
			T.~S.~Bunch and L.~Parker,
			``Feynman Propagator in Curved Space-Time: A Momentum Space Representation,''
			Phys.\ Rev.\ D {\bf 20}, 2499 (1979).
			%doi:10.1103/PhysRevD.20.2499
			%%CITATION = doi:10.1103/PhysRevD.20.2499;%%
			%217 citations counted in INSPIRE as of 1
			
			%\cite{Bunch:1980vc}
			\bibitem{Bunch:1980vc}
			T.~S.~Bunch,
			``Adiabatic Regularization For Scalar Fields With Arbitrary Coupling To The Scalar Curvature,''
			J.\ Phys.\ A {\bf 13}, 1297 (1980).
			%doi:10.1088/0305-4470/13/4/022
			%%CITATION = doi:10.1088/0305-4470/13/4/022;%%
			%109 citations counted in INSPIRE as of 14 Nov 2018
			
				%\cite{Alberte:2015cxa}
			\bibitem{Alberte:2015cxa}
			L.~Alberte, R.~Brustein, A.~Khmelnitsky and A.~J.~M.~Medved,
			``Density matrix of black hole radiation,''
			JHEP {\bf 1508}, 015 (2015)
			%doi:10.1007/JHEP08(2015)015
			[arXiv:1502.02687 [hep-th]].
			

			
			%\cite{Poisson:2004cw}
			\bibitem{Poisson:2004cw}
			E.~Poisson,
			``Absorption of mass and angular momentum by a black hole: Time-domain formalisms for gravitational perturbations, and the small-hole / slow-motion approximation,''
			Phys.\ Rev.\ D {\bf 70}, 084044 (2004)
			%	doi:10.1103/PhysRevD.70.084044
			[gr-qc/0407050].
			%%CITATION = doi:10.1103/PhysRevD.70.084044;%%
			%77 citations counted in INSPIRE as of 29 May 2018
			
			%\cite{Poisson:2009qj}
			\bibitem{Poisson:2009qj}
			E.~Poisson and I.~Vlasov,
			``Geometry and dynamics of a tidally deformed black hole,''
			Phys.\ Rev.\ D {\bf 81}, 024029 (2010)
			%	doi:10.1103/PhysRevD.81.024029
			[arXiv:0910.4311 [gr-qc]].
			
			
			%\cite{Taylor:2008xy}
			\bibitem{Taylor:2008xy}
			S.~Taylor and E.~Poisson,
			``Nonrotating black hole in a post-Newtonian tidal environment,''
			Phys.\ Rev.\ D {\bf 78}, 084016 (2008)
			%doi:10.1103/PhysRevD.78.084016
			[arXiv:0806.3052 [gr-qc]].
			%%CITATION = doi:10.1103/PhysRevD.78.084016;%%
			%53 citations counted in INSPIRE as of 18 Dec 2018
			
			
			%\cite{Poisson:2018qqd}
			\bibitem{Poisson:2018qqd} 
			E.~Poisson and E.~Corrigan,
			``Nonrotating black hole in a post-Newtonian tidal environment II,''
			Phys.\ Rev.\ D {\bf 97}, no. 12, 124048 (2018)
			%	doi:10.1103/PhysRevD.97.124048
			[arXiv:1804.01848 [gr-qc]].
			%%CITATION = doi:10.1103/PhysRevD.97.124048;%%{{}
			
			%\cite{Thorne:1980ru}
			\bibitem{Thorne:1980ru}
			K.~S.~Thorne,
			``Multipole Expansions of Gravitational Radiation,''
			Rev.\ Mod.\ Phys.\  {\bf 52}, 299 (1980).
			%doi:10.1103/RevModPhys.52.299
			%%CITATION = doi:10.1103/RevModPhys.52.299;%%
			%748 citations counted in INSPIRE as of 19 Dec 2018
			
					
			%\cite{Vega:2011ue}
			\bibitem{Vega:2011ue}
			I.~Vega, E.~Poisson and R.~Massey,
			``Intrinsic and extrinsic geometries of a tidally deformed black hole,''
			Class.\ Quant.\ Grav.\  {\bf 28}, 175006 (2011)
			%doi:10.1088/0264-9381/28/17/175006
			[arXiv:1106.0510 [gr-qc]].
			%%CITATION = doi:10.1088/0264-9381/28/17/175006;%%
			%20 citations counted in INSPIRE as of 06 Sep 2018	
			
			
			
					%\cite{Price:1994pm}
			\bibitem{Price:1994pm} 
			R.~H.~Price and J.~Pullin,
			``Colliding black holes: The Close limit,''
			Phys.\ Rev.\ Lett.\  {\bf 72}, 3297 (1994)
			%	doi:10.1103/PhysRevLett.72.3297
			[gr-qc/9402039].
			%%CITATION = doi:10.1103/PhysRevLett.72.3297;%%
			%186 citations counted in INSPIRE as of 09 Jul 2019
			
			
			
	
			
			
			
			
		
			
			
		
			
			
			
			
			
			

			
					

		
			
	
	
		
			
			
			
			
			
			
			
			
			
			
			
			
			
			

			
			
			
			
			
			
			
			
			
			
		
			
	
			
			

			
			
			
			
		}
		
		
		
		
		
		
		
	\end{thebibliography}
\end{document}